# Response to comment on "Broken translational and rotational symmetry via charge stripe order in underdoped YBa$_2$Cu$_3$O$_{6+y}$"

**Authors:** R. Comin,[1,2,*] R. Sutarto,[3] E. H. da Silva Neto,[1,2,4,5] L. Chauviere,[1,2,5] R. Liang,[1,2] W. N. Hardy,[1,2] D. A. Bonn,[1,2] F. He,[3] G. A. Sawatzky,[1,2] A. Damascelli[1,2*]

**Affiliations:**

[1] Department of Physics and Astronomy, University of British Columbia, Vancouver, British Columbia V6T 1Z1, Canada.

[2] Quantum Matter Institute, University of British Columbia, Vancouver, British Columbia V6T 1Z4, Canada.

[3] Canadian Light Source, Saskatoon, Saskatchewan S7N 2V3, Canada.

[4] Quantum Materials Program, Canadian Institute for Advanced Research, Toronto, Ontario M5G 1Z8, Canada.

[5] Max Planck Institute for Solid State Research, D-70569 Stuttgart, Germany.

* Email: r.comin@utoronto.ca; damascelli@physics.ubc.ca

**Abstract**: Fine questions our interpretation of unidirectional-stripes over bidirectional-checkerboard, and illustrates his criticism by simulating a momentum space structure consistent with our data and corresponding to a checkerboard-looking real space density. Here we use a local rotational-symmetry analysis to demonstrate that the simulated image is in actuality composed of locally unidirectional modulations of the charge density, consistent with our original conclusions.

**Main Text:**

We recently found that the symmetry of the underlying instability responsible for the charge-order observed in the cuprate superconductor YBa$_2$Cu$_3$O$_{6+y}$ (YBCO) is of a unidirectional (1D) stripe rather than bidirectional (2D) checkerboard type (1). Fine (2) questions this conclusion by providing an example of charge modulation seemingly consistent with a bidirectional checkerboard modulation, but having the necessary characteristics of one-dimensional order as per the definitions in (1). In the following, we will show that this is not the case.

The charge modulation patterns presented in figure 1, B and C, of Fine's comment give the overall impression of a globally 2D order. However, establishing whether this corresponds to a true checkerboard or stems from the coexistence of two independent 1D modulations is a much subtler issue, even in the simple case of Fine's Fig. 1C. A unidirectional charge order instability is accompanied by the breaking of fourfold symmetry, but its detection can be obscured by the presence of 90° rotated domains giving rise to what might appear as a globally 2D modulation. We have analyzed the density wave pattern in Fine's figure 1C and determined that it locally breaks $D_{4h}$ symmetry and is thus best described as the superposition of two independent, locally-1D modulations.

In (1) we implemented an analysis based on these symmetry considerations, which allows discriminating the two scenarios discussed above. We note that our conclusion in favor of charge-stripe order is supported by two completely independent pieces of evidence: (i) the symmetry analysis of the RXS structure factor performed for a variety of configurations, including domains as well as canted domains (this is what Fine questions in his comment); (ii) the intrinsic unidirectionality observed for the competition between charge order and superconductivity [see figure 3 in (1) and the discussion of the strong anisotropy in the temperature evolution of the charge-order correlation length].

Hereafter we will present quantitative counterarguments to the thesis outlined in Fine's comment. Stripe order is a smectic charge ordered state (3), i.e. one breaking both translational and fourfold rotational ($D_{4h}$) symmetry. In order to reveal and characterize the local breaking of these symmetries, we have analyzed the real-space density map proposed by Fine in his figure 1C in terms of its local $D_{4h}$-symmetry breaking. We seek to isolate the two real-space density components that are modulated along x and y [$\delta\rho_x(\mathbf{r})$ and $\delta\rho_y(\mathbf{r})$]. We do this by taking a local Fourier transform that is often used to analyze scanning tunneling microscopy data [see e.g. the Supplementary materials of (4)]:

$$\delta\rho_x(\mathbf{r}) = \frac{1}{2\pi\Gamma^2}\sum_{\mathbf{r}'} \delta\rho(\mathbf{r}') \cdot e^{i\mathbf{Q}_x \cdot \mathbf{r}'} \cdot e^{-\frac{|\mathbf{r}'-\mathbf{r}|^2}{2\Gamma^2}},$$

and similarly for $\delta\rho_y(\mathbf{r})$. Here $\delta\rho(\mathbf{r})$ is the original real-space density pattern, and $\Gamma$ is chosen to be appropriate to the domain sizes observable in $\delta\rho(\mathbf{r})$ ($\Gamma \sim 0.22\ Q^{-1}$). As a formally defined metric for the checkerboard state, we use two similar definitions of a local order parameter, following a previous theoretical study (5) and more recent experimental STM work (6). These two quantities, respectively denominated $\Sigma_1(\mathbf{r})$ and $\Sigma_2(\mathbf{r})$, are defined as follows:

$$\Sigma_1(\mathbf{r}) = \left(|\delta\rho_y(\mathbf{r})|^2 - |\delta\rho_x(\mathbf{r})|^2\right) / \left(|\delta\rho_y(\mathbf{r})|^2 + |\delta\rho_x(\mathbf{r})|^2\right)$$

$$\Sigma_2(\mathbf{r}) = \left(|\delta\rho_y(\mathbf{r})| - |\delta\rho_x(\mathbf{r})|\right) / \left(|\delta\rho_y(\mathbf{r})| + |\delta\rho_x(\mathbf{r})|\right)$$

These local order parameters are bound between -1 and 1; whereas +1 (-1) corresponds to stripes propagating along y (x), 0 is a pure checkerboard state (i.e., fourfold symmetric) where the two density modulations need to have locally the same amplitude, a necessary condition for $D_{4h}$ point group symmetry to be preserved.

The plot of $\Sigma_1(\mathbf{r})$ is superimposed on the real-space density map (Fig. 1, A and B). The presence of multiple colorful patches implies the existence of extended regions with local stripe modulations predominantly along y [i.e., $\Sigma_1(\mathbf{r}) = 1$, red] or predominantly along x [i.e., $\Sigma_1(\mathbf{r}) = -1$, blue]. The choice of a threshold value for $\Sigma_1(\mathbf{r})$ separating bidirectional from unidirectional nanoregions would be arbitrary, and ultimately only the extremal values correspond to well-defined order parameters. However, the existence of extended stripy regions reveals a clear tendency of the charge density to locally break $D_{4h}$ point group symmetry. Similar observations can be made for the corresponding plots of $\Sigma_2(\mathbf{r})$ (Fig 1, C and D), where blue and red ellipses in

Fig. 1C illustrate the location of x and y stripy domains, respectively, confirming the existence of extended regions with local stripe modulations.

We note that even in the original density pattern it can be observed that most regions approach a unidirectional character, even though the general appearance of the image gives an overall impression of a checkerboard motif. This is visualized in the x- and y-projected density maps of Fig. 1E [$\delta\rho_x(\mathbf{r})$] and 1F [$\delta\rho_y(\mathbf{r})$].

The overall picture is essentially what one would obtain by taking the real-space schematics in figure 2A of (1) and partially overlapping the stripy domains on top of each other. An equivalent configuration (from the perspective of the reciprocal space representation) would be obtained with a criss-cross pattern of stripes in adjacent $CuO_2$ planes within a bilayer, because of our inability to resolve the c-axis-projected structure of the charge-ordered state. In (1) we showed only disjointed stripy domains for illustrative purposes; we did not assume that the individual stripy domains have to be segregated.

Finally, we can also calculate the cross-correlation coefficient $r_{\rho_x,\rho_y} = \frac{1}{n}\sum_\mathbf{r} \frac{(\rho_x(\mathbf{r})-\overline{\rho_x})\cdot(\rho_y(\mathbf{r})-\overline{\rho_y})}{\sigma_{\rho_x}\cdot\sigma_{\rho_y}}$, where $\rho_x$ ($\rho_y$) is the density amplitudes along x (y), $\sigma_{\rho_x}$ ($\sigma_{\rho_y}$) the corresponding standard deviation, and $n$ is the number of discretized spatial points. This leads to $r_{\rho_x,\rho_y} \sim 0.25$, once again closer to the limit of pure stripe ($r_{\rho_x,\rho_y} = 0$) than checkerboard ($r_{\rho_x,\rho_y} = 1$) order.

In conclusion, the specific counterexample provided by Fine also belongs to the general case of a locally unidirectional charge modulation, albeit with partially overlapping 90°-rotated domains, as opposed to an actual checkerboard. Such a topology is more straightforward to distinguish from a native checkerboard in reciprocal space than it is in real space. Our analysis here highlights the compatibility of an overlapping stripe-domains' configuration with the momentum-space structure determined experimentally in our study. Finally, this also reaffirms our conclusion in favor of a microscopic symmetry breaking via charge-stripe order, consistent with the strong anisotropy observed for the temperature evolution of the correlation length.

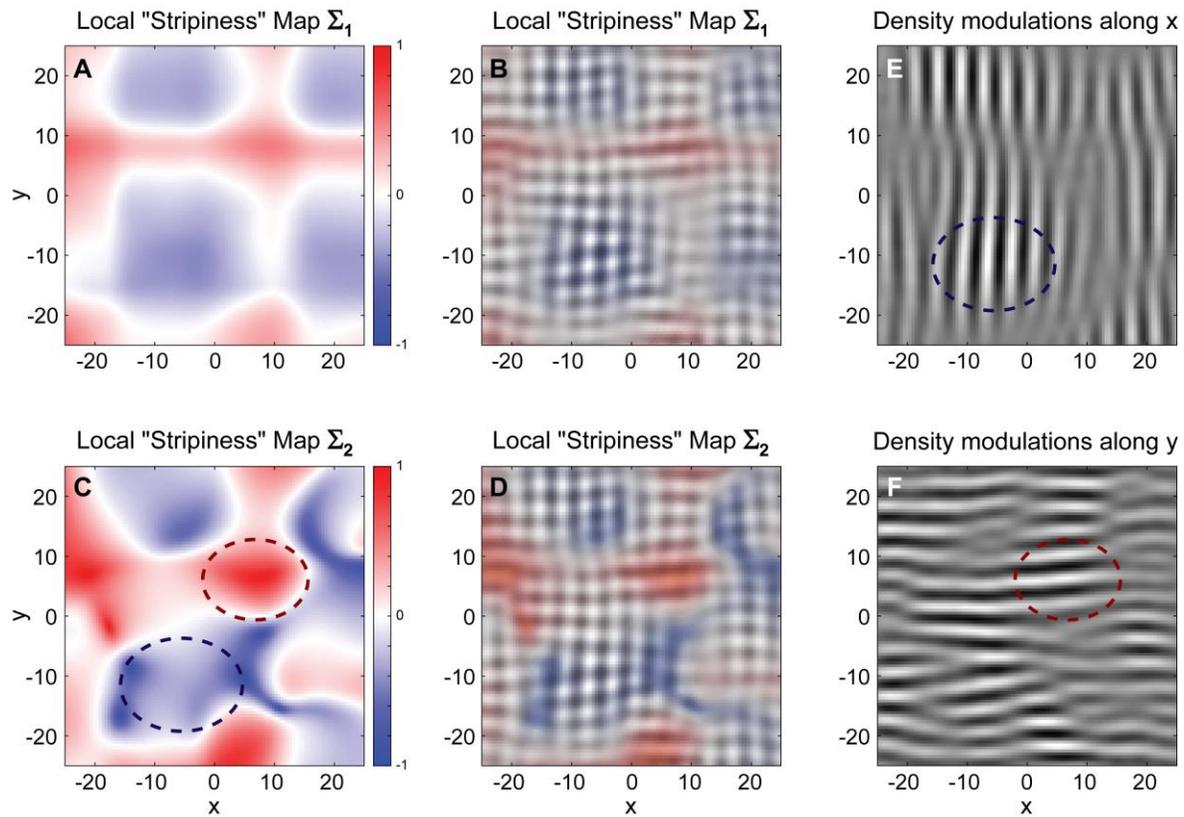

**Fig. 1. Symmetry analysis in real and reciprocal space.** (**A**) Map of the order parameter (local "stripiness") $\Sigma_1(\mathbf{r})$ calculated as outlined in the text, overlaid on top of the original map in (**B**). (**C** and **D**) Same as (A) and (B), but for the $\Sigma_2(\mathbf{r})$ order parameter. Density modulations along x- (**E**) and y (**F**), extracted from the dataset used in figure 1C of Fine's comment [blue and red ellipses are the same as in (C)].